\documentclass[%
 prl,
 twocolumn,
nofootinbib,
 amsmath,amssymb,
 aps,
preprintnumbers,
superscriptaddress
]{revtex4-2}

\usepackage{graphicx}% Include figure files
\usepackage{dcolumn}% Align table columns on decimal point
\usepackage{bm}% bold math
\usepackage{hyperref}% add hypertext capabilities
\usepackage{color}
\usepackage[normalem]{ulem}

\newcommand{\GeV}{\rm GeV}

\begin{document}

\preprint{PITT-PACC-2119}

\title{Thermal Misalignment of Scalar Dark Matter}

\author{Brian Batell}%
 \email{batell@pitt.edu}
\author{Akshay Ghalsasi}
 \email{akg53@pitt.edu}

\affiliation{%
Pittsburgh Particle Physics, Astrophysics, and Cosmology Center, Department of Physics and Astronomy,
University of Pittsburgh, Pittsburgh, USA
}%

\date{\today}% It is always \today, today,

\begin{abstract}
The conventional misalignment mechanism for scalar dark matter depends on the initial field value, which governs the oscillation amplitude and present-day abundance. We present a mechanism by which a feeble (Planck-suppressed) coupling of dark matter to a fermion in thermal equilibrium drives the scalar towards its high-temperature potential minimum at large field values, dynamically generating misalignment before oscillations begin. Unlike conventional misalignment production, the dark matter abundance is dictated by microphysics and not by initial conditions. As an application of the generic mechanism, we discuss a realistic scenario in which dark matter couples to the muon. 
\end{abstract}

\maketitle

%\tableofcontents

{\bf Introduction.}---There is by now overwhelming evidence for the existence of dark matter (DM), which makes up about a quarter of the energy budget of our universe~\cite{Planck:2018vyg}, but many open questions about its fundamental nature persist. Among the most basic of these are the underlying particle physics dynamics of DM and its genesis in the early universe. 
In a well-motivated and widely studied class of models, ultralight scalar bosonic DM $\phi$ with 
mass $10^{-22} \, {\rm eV} \lesssim m_\phi \lesssim \, {\rm keV}$ is generically produced in the early universe through the {\it misalignment mechanism}~\cite{Preskill:1982cy,Abbott:1982af,Dine:1982ah}. Starting from some initial field value $\phi_i$ at some early time $t_i$, the scalar field begins to oscillate once the Hubble expansion rate falls below its mass and subsequently behaves as cold DM (i.e., its mean energy density scales with the inverse cube of the cosmic scale factor, it has vanishing mean pressure,
etc.). 

In the conventional misalignment mechanism just described, the late time oscillation amplitude and resulting abundance depends on the initial field value $\phi_i$. Unlike other popular DM production scenarios, such as thermal freeze-out of weak-scale DM, the abundance is not solely governed by fundamental particle physics parameters such as masses and interaction strengths but is sensitive to initial conditions. In this \textit{Letter}, we present a simple and generic mechanism to dynamically generate large scalar DM misalignment starting from fairly generic initial conditions. The mechanism relies on a finite temperature scalar potential generated by a coupling to a fermion in the thermal bath, which drives the scalar field towards its high temperature minimum at large field values, thereby dynamically generating misalignment.
Provided the initial field value is small in comparison to the eventual oscillation amplitude, the present-day abundance is completely determined by the DM microphysics and is insensitive to the precise initial conditions.  

Because of the simplicity of the setup, the thermal misalignment mechanism can easily be realized in a variety of realistic particle physics models. 
In addition, since the mechanism relies on the coupling of DM to a fermion, there are in general novel phenomenological opportunities to probe DM in comparison to the conventional misalignment mechanism. As an illustration, below we examine one realistic scenario in which the scalar DM couples to the muon. This scenario features a rich variety of observational and experimental probes that can test regions of parameter space explaining the observed DM abundance. 
We note that modified scalar dynamics due to thermal effects or novel interactions has been considered in other contexts, such as %context of  
mass varying neutrinos~\cite{Fardon:2003eh,Fardon:2005wc,Weiner:2005ac,Ghalsasi:2016pcj}, scalar trapping \cite{Moroi:2013tea} and axions \cite{Brzeminski:2020uhm, DiLuzio:2021gos,Co:2018mho,Co:2018phi}. 

{\bf Minimal model and mechanism.}---The basic model realizing the dynamical misalignment mechanism consists of a real scalar DM field $\phi$ and a Dirac fermion $\psi$, with Lagrangian
\begin{equation}
\label{eq:Lagrangian}
-{\cal L} 
= \frac{1}{2} \,m_\phi^2 \, \phi^2 + m_\psi \, \left(1-\frac{\beta\,\phi}{M_{\rm pl}} \right)\overline \psi \,\psi,
\end{equation}
where $m_\phi$ ($m_\psi$) is the scalar (fermion mass) and $M_{\rm pl} = (8 \pi G_N)^{-1/2} = 2.4 \times 10^{18}$ GeV is the reduced Planck mass.
The fields interact through a Yukawa coupling, which for later convenience we have parameterized as $-\beta m_\psi/M_{\rm pl}$ with $\beta$ a real dimensionless parameter.  

\begin{figure}
\includegraphics[width=\linewidth]{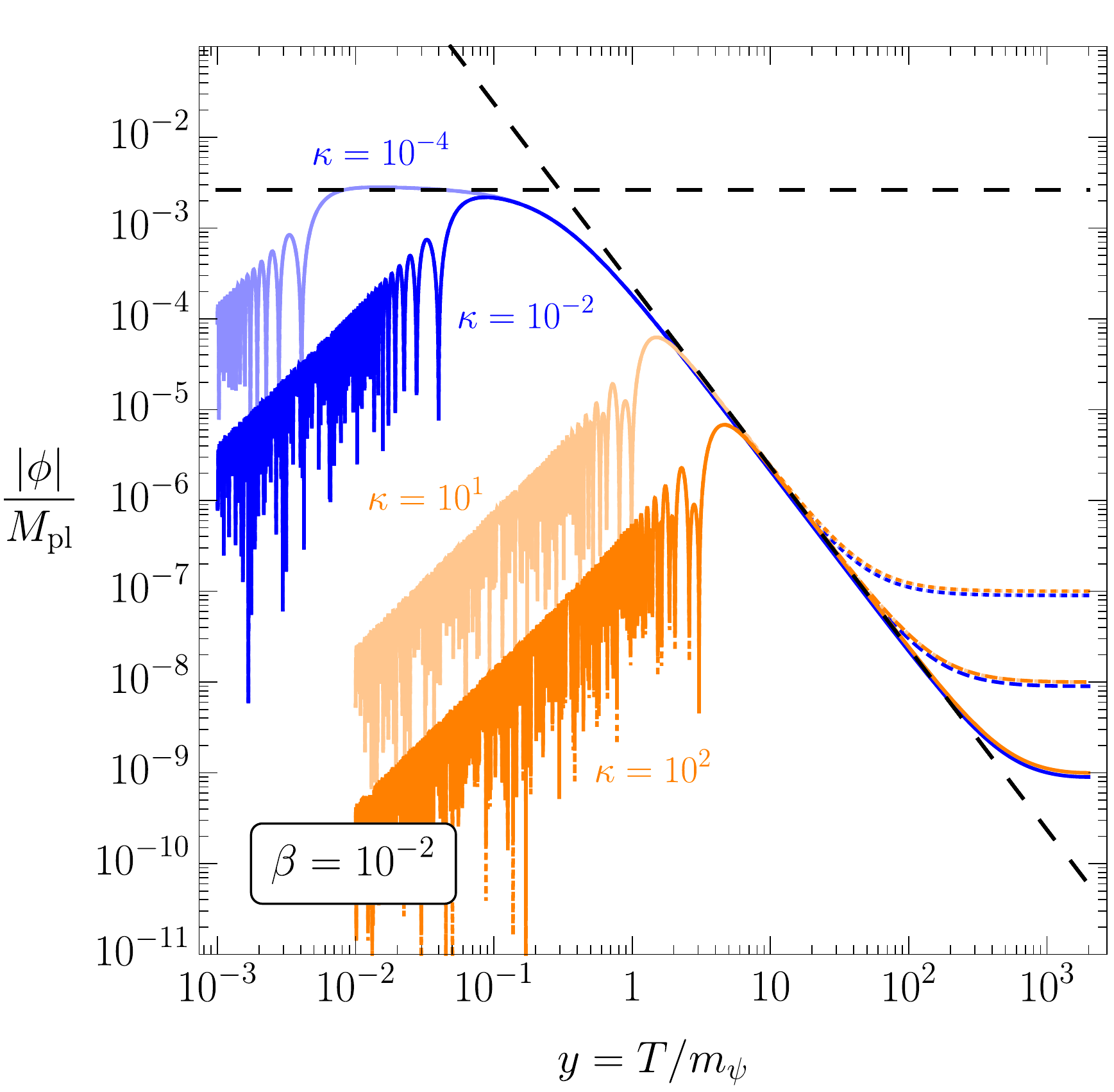}~
\caption{Scalar field evolution for $\beta = 10^{-2}$ and $\kappa = [10^{-4},10^{-2},10,100]$ where $\kappa \equiv m_\phi M_{\rm pl}/ m_\psi^2$ for $y_{\rm osc} > 1$ (orange lines, Region 1) and $y_{\rm osc} < 1$ (blue lines, Region 2). Dashed black lines show the analytical approximations  
of Eqs.~(\ref{eq:phi-sol-initial-t},\ref{eq:R2-phi-osc}).  The final yield is independent of the initial value $\phi_{i}$ of the scalar field. We have assumed constant $g_{*} = 10.75$ throughout the evolution.
\label{fig:scalar-evolution}
}
\end{figure}

The envisioned cosmological history is as follows. We assume the fermion $\psi$ attains thermal equilibrium with the SM radiation bath in the early universe. The scalar $\phi$ acquires a time dependent, spatially homogeneous background field value, which evolves according to the equation of motion
\begin{equation}
\ddot{\phi} + 3 H \dot{\phi} + \frac{dV_{\rm eff}}{d\phi}=0.
\label{eq:EOM}
\end{equation}
Here $H = 1/2t = \gamma T^2/M_{\rm pl}$ is the Hubble parameter in the radiation era, where $t$ denotes time and $\gamma(T) = \sqrt{\pi^2 g_*(T)/90}$ with $g_{*(S)}(T)$ the effective number of relativistic (entropy) degrees of freedom. It will often be convenient use the variable 
$y \equiv T/m_{\psi}$. When $y \gg 1$ ($y\ll 1$) the fermions are in thermal equilibrium (Boltzmann suppressed).
The scalar effective potential $V_{\rm eff}$ appearing in Eq.~(\ref{eq:EOM}) includes the tree level contribution from Eq.~(\ref{eq:Lagrangian}) along with a finite-temperature correction arising from the thermal free energy density of $\psi$~\cite{Dolan:1973qd,Weinberg:1974hy},
\begin{align}
\label{eq:Veff1}
\delta V_{T}(\phi)& =  -\frac{g_\psi}{2\pi^{2}}T^{4} J_F \left[
\frac{m^2_\psi(\phi)}{T^2}
\right],
\end{align}
where $g_\psi = 4$ counts the fermion spin degrees of freedom, $m_\psi(\phi) = m_\psi(1-\beta \phi/M_{\rm pl})$ 
is the effective fermion mass in the scalar background, and 
\begin{align}
\label{eq:JF}
J_F(w^2) = \int_0^\infty \!\!\! dx \,x^2 \, \log\left[\,1+e^{-\displaystyle\sqrt{x^2+w^2}}\,\right].
\end{align}
The correction to the effective potential~(\ref{eq:Veff1}) leads to the development of a high-temperature minimum at large scalar field values. The scalar will then evolve from generic small initial field values towards the high temperature minimum, generating misalignment. As the temperature drops and the Hubble rate falls below the effective scalar mass, $\phi$ begins to oscillate at some temperature $T_{\rm osc}$ ($y_{\rm osc} \equiv T_{\rm osc}/m_\psi$) and ultimately behaves as DM.
In Fig.~\ref{fig:scalar-evolution} we display the numerical evolution of $\phi$ with $y$ for several choices of model parameters and initial conditions, illustrating the generation of large scalar misalignment from generic small initial field values and the subsequent late-time oscillations. 

The general features of the thermal misalignment mechanism just outlined are most easily understood through an analysis of the dynamics at high temperatures, $T\gg m_\psi(\phi)$.
The scalar effective potential in this regime, including the zero temperature quadratic term (\ref{eq:Lagrangian}) and the thermal free energy density 
($\phi$-dependent terms) 
(\ref{eq:Veff1}), is given by
\begin{align}
\label{eq:Veffzg1}
V_{\rm eff} \simeq  \frac{1}{2}m^{2}_{\phi} \phi^{2} +  \frac{T^{2}m^2_\psi}{12}\left(1 -\frac{\beta \phi }{M_{\rm pl}}\right)^{2} .
\end{align}
The minimum of this potential is
\begin{align}
\label{eq:Amin}
\phi_{\rm min}|_{T \gg m_\psi(\phi)} = \frac{\beta m_\psi^{2} M_{\rm pl} T^{2}}{6 m^{2}_{\phi} M^{2}_{\rm pl}+ m_\psi^{2}T^{2}\beta^{2}} = M_{\rm pl} \frac{\beta y^{2}}{\beta^{2} y^{2} +6\kappa^{2}},
\end{align}
where $y$ is defined below Eq.~(\ref{eq:EOM}) and we have introduced the dimensionless parameter $\kappa \equiv m_\phi M_{\rm pl}/ m_\psi^2$.
The potential minimum (\ref{eq:Amin}) results from the competition between the linear and quadratic terms in the effective potential~(\ref{eq:Veffzg1}). We see that at very high temperatures, $y\gg\sqrt{6}\kappa/\beta$, the second term in (\ref{eq:Veffzg1}) dominates and the minimum is located at the large field value $\phi_{\rm min} \simeq M_{\rm pl}/\beta$. At somewhat lower temperatures  $1 \leq y\ll\sqrt{6}\kappa/\beta$, the quadratic term is dominated by the first term in (\ref{eq:Veffzg1}), and the minimum is located at $\phi_{\rm min} \simeq M_{\rm pl} \beta y^2 /6 \kappa^2$. In the very low temperature regime, $y \ll 1$, the fermions are Boltzmann suppressed, 
$\delta V_T \propto e^{-m_{\psi}(\phi)/T}$, and the minimum moves toward the origin. 

During the initial stages of the evolution, the effective potential (\ref{eq:Veffzg1}) is dominated by the linear term, $V_{\rm eff} \supset - T^{2} m_\psi^{2} \beta  \phi /6 M_{\rm pl}$, and the scalar satisfies the condition, $|\ddot \phi| \ll |H \dot \phi|$. Therefore, the equation of motion (\ref{eq:EOM}) simplifies dramatically, 
\begin{equation}
\dot \phi \simeq \frac{\beta m_\psi^2}{18 \gamma}.
\end{equation}
Neglecting the mild variation of $g_*$ with temperature and integrating this equation, we obtain  
\begin{align}
\label{eq:phi-sol-initial-t}
\phi(t) & = \phi_i\! +\! \frac{\beta m_\psi^2}{18 \gamma} (t\! - \! t_i) \simeq \frac{\beta m_\psi^2}{18 \gamma} t ~~\,\rightarrow ~~\,
\phi(y) \simeq \frac{\beta M_{\rm pl}}{36 \gamma^2 } \frac{1}{y^2}.
\end{align}
Provided the initial value of the field is smaller than its eventual value at the onset of scalar oscillations, $|\phi_{i}| \ll \phi_{\rm osc} \equiv \phi(y_{\rm osc})$ and $\phi_{\rm osc} \ll M_{\rm pl}/\beta$, we observe that the approximate early-time solution (\ref{eq:phi-sol-initial-t}) is not sensitive to the initial conditions and grows in proportion to the cosmic time, generating misalignment. This behavior is also apparent from numerical solution shown in Fig.~\ref{fig:scalar-evolution}. 
Below we will use the early-time solution (\ref{eq:phi-sol-initial-t}) as input in our estimates of $\phi_{\rm osc}$.

\begin{figure}
\includegraphics[width=\linewidth]{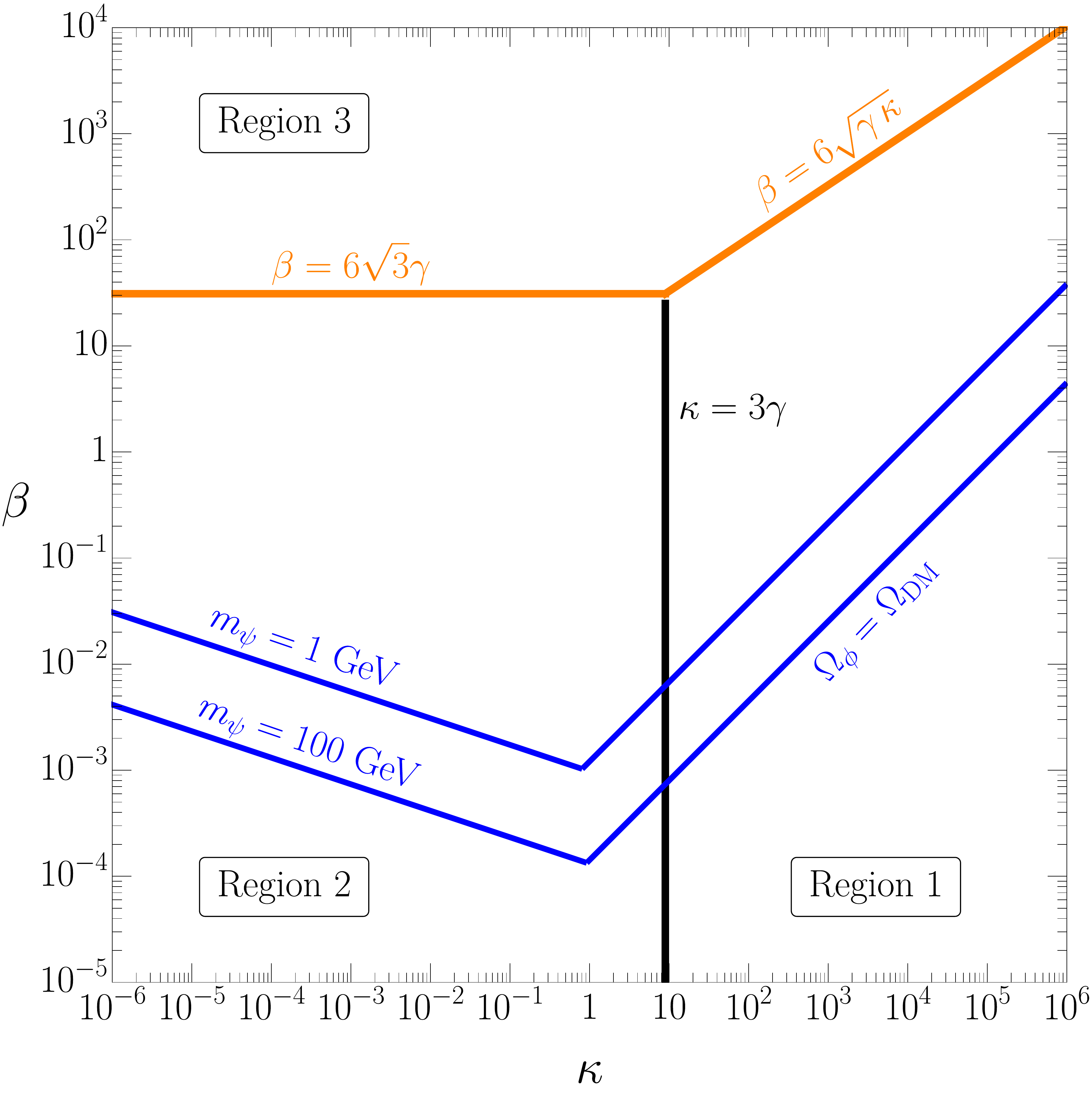}~
\caption{Regions 1,2,3 in the $\kappa-\beta$ plane. Parameters predicting the correct DM abundance, $\Omega_{\phi} = \Omega_{\rm DM}$, are indicated by the blue lines for $m_{\psi} = 1\, \GeV, 100 \,\GeV $, where for simplicity we have assumed $g_{*} = g_{*}(m_{\psi})$. The boundaries defining the three regions are drawn for $g_{*}(T = 1\, {\rm GeV}) \simeq 81$. \label{fig:pspace}
}
\end{figure}
As the universe expands and the temperature drops, the expansion rate eventually becomes smaller than the effective scalar mass, signaling the beginning of scalar oscillations. 
From Eq.~(\ref{eq:Veffzg1}) we obtain the effective scalar mass at high temperatures, 
\begin{align}
\label{eq:meff}
m^{2}_{\phi}(T) = m^{2}_{\phi} + \frac{ \beta^{2}  m_\psi^{2}  T^{2}   }{6 M^{2}_{\rm pl}} = m^{2}_{\phi}\left(1 +  \frac{\beta^{2} \, y^{2}      }{6\, \kappa^{2}}\right).
\end{align}
Considering that the oscillations begin for $3H(T_{\rm osc}) = m_{\phi}(T_{\rm osc})$ the oscillation temperature is estimated as 
\begin{align}
\label{eq:yosc}
y_{\rm osc} = \frac{\beta}{6\sqrt{3}\gamma}\sqrt{\left(1 +\sqrt{1+\frac{1296 \gamma^{2} \kappa^{2}}{\beta^{4}}}\right)} \,.
\end{align}
For $y_{\rm osc} \geq 1$ and $\beta \gg 6\sqrt{\gamma \kappa}$, 
the oscillations begin at $y_{\rm osc}\simeq \beta/6\sqrt{3}\gamma$. 
However, for $\beta \ll 6\sqrt{\gamma \kappa}$, $y_{\rm osc} \simeq \sqrt{\kappa/ 3\gamma}$ and is controlled by the zero temperature mass of the scalar. This motivates a division of the $\kappa-\beta$ parameter space into three regions, with boundaries defined by the conditions 
$y_{\rm osc} = 1$ and $\beta = 6 \sqrt{\gamma \kappa}$, as shown in Fig.~\ref{fig:pspace}.
We now study both Regions 1 and 2, where the scalar begins oscillating under its zero temperature mass (below the orange line in Fig.~\ref{fig:pspace}), in order to obtain an analytical understanding of the $\phi$ evolution and the eventual DM yield.  A detailed description of Region 3 (above the orange line) will be presented elsewhere.

We first discuss Region 1, which is 
defined by 
\begin{equation}
\label{eq:R1}
y_{\rm osc} \simeq \sqrt{\frac{\kappa}{3\gamma}} > 1 ~~ (\kappa > 3\gamma) ~~~~
 {\rm and} ~~~~ \beta < 6 \sqrt{\gamma\,\kappa}.
\end{equation}
In this region the scalar oscillations are primarily controlled by their zero temperature mass and begin before the fermions 
leave the plasma. 
So even though the fermions do not control the onset of oscillations, the amplitude of the oscillations is dictated by the scalar-fermion coupling $\beta$. An estimate of the field value $\phi_{\rm osc}$  at the beginning of oscillations is obtained by evaluating Eq.~(\ref{eq:phi-sol-initial-t}) at $y = y_{\rm osc} \simeq \sqrt{\kappa/ 3\gamma}$:
\begin{equation}
\label{eq:R1-phi-osc}
\phi_{\rm osc} \equiv \phi(y_{\rm osc}) \simeq \frac{\beta M_{\rm {\rm pl}}}{12 \gamma \kappa}.
\end{equation}
The present-day DM energy density is given by  
$\rho_{\phi,0} = \displaystyle\tfrac{1}{2}m^{2}_{\phi} \phi^{2}_{\rm osc} (y_{\rm 0}/y_{\rm osc})^3(g^{\rm 0}_{*S}/g^{\rm osc}_{*S})$, where $y_0 = T_0 / m_\psi$ with $T_0 = 2.7$ K and $g_{*S}^0 \simeq 3.91$. 
Using this result and Eqs.~(\ref{eq:R1},\ref{eq:R1-phi-osc}), the DM density parameter today, $\Omega_{\phi} \equiv \rho_{\phi,0}/\rho_{c,0}$ with $\rho_{c,0}=3M^2_{\rm pl}H_0^2$ the critical density, is estimated as
\begin{align}
\label{eq:Omega-Phi-R1}
\Omega_{\phi} & \simeq      \Omega_{\rm DM}   \left(\frac{m_\psi}{0.1 \, \rm GeV}\right)  \left(\frac{\beta}{0.1 }\right)^2  \left(\frac{400}{\kappa}\right)^{3/2} \left(\frac{10.75}{g^{\rm osc}_{*S}}\right)^{5/4},
\end{align}
where $\Omega_{\rm DM} \simeq 0.26$~\cite{Planck:2018vyg}.
We next consider Region 2, which is 
defined by 
\begin{equation}
\label{eq:R2}
y_{\rm osc} \simeq \sqrt{\frac{\kappa}{3\gamma}} < 1 ~~ (\kappa < 3\gamma) ~~~~
 {\rm and} ~~~~ \beta < 6 \sqrt{3} \gamma .
\end{equation}
In this region the oscillations begin after the fermions are Boltzmann suppressed and no longer affect the evolution of $\phi$. So until $y\sim 1$ the solution is given by Eq.~(\ref{eq:phi-sol-initial-t}), $\phi \sim \beta M_{\rm pl}/y^2$.
Then, for $y_{\rm osc} \ll 1$, 
the velocity of $\phi$ experiences Hubble friction and reaches an asymptotic value of  
\begin{equation}
\label{eq:R2-phi-osc}
\phi_{\rm osc} \simeq 0.27 \frac{ \beta M_{\rm pl}}{\gamma^{2}}
\end{equation}
before oscillations start. Similarly to Region 1, we can estimate the dark matter density parameter today:
\begin{align}
\label{eq:Omega-Phi-R2}
\Omega_{\phi} & \simeq       \Omega_{\rm DM}    \left(\frac{m_\psi}{0.1 \, \rm GeV}\right)  \left(\frac{\beta}{10^{-3} }\right)^2  \left(\frac{\kappa}{0.01}\right)^{1/2} \left(\frac{10.75}{g^{\rm osc}_{*S}}\right)^{9/4}.
\end{align}
This shows that in both regions the DM abundance depends mainly on the coupling $\beta$ and the DM mass $m_\phi$. 

In Fig.~\ref{fig:pspace} we show the parameter choices where Eqs.~(\ref{eq:Omega-Phi-R1},\ref{eq:Omega-Phi-R2}) predict the observed DM abundance. Near $\kappa \sim 3 \gamma$, the transition between Regions 1 and 2, we have extrapolated these predictions to their intersection. In our phenomenological example below, we will compare this with abundance prediction from the exact numerical evolution of the system. Fig.~\ref{fig:pspace} shows that the correct DM abundance can be obtained over a broad range of masses and couplings. 

Before examining a realistic scenario in which the fermion is the muon, a few remarks are in order. 
First, we note that along with the finite temperature correction~(\ref{eq:Veff1}), the effective potential receives a zero temperature correction at one-loop, i.e., the Coleman-Weinberg potential~\cite{Coleman:1973jx}.
We assume here that the full zero temperature effective potential is well described by a simple quadratic potential as in Eq.~(\ref{eq:Lagrangian}). This implies the mass term as well as quartic coupling $\lambda$ are fine-tuned for small scalar masses $m_\phi$ and large couplings $\beta$. This is a manifestation of the well-known naturalness problem associated with light scalars. In our phenomenological example below, we will indicate regions of parameter space where such fine-tuning is needed. Though beyond our present scope, it would be worthwhile to explore model building avenues to protect such light, weakly coupled scalars; see for example Refs.~\cite{Hook:2018jle,Brzeminski:2020uhm} for recent promising work in this direction.  

The inflationary epoch  can potentially impact the thermal misalignment production mechanism. The classical and quantum evolution of $\phi$ during inflation leads to a characteristic range of field values at the end of inflation, which should be compared with the requirement on our initial conditions described above, $\phi_{i} \ll \phi(y_{\rm osc})$. Moreover, the scalar fluctuations at the end of inflation contribute to isocurvature perturbations, which are strongly constrained by CMB data \cite{Planck:2018jri}. However, assuming a long enough inflationary period (which relaxes the scalar to its zero temperature minimum) with a low enough Hubble scale during inflation (which suppresses the scalar fluctuations), we can avoid both isocurvature constraints and a fine-tuning of our scalar field initial conditions~\cite{Tenkanen:2019aij,Graham:2018jyp}. 

\begin{figure*}
    \includegraphics[width=0.9\linewidth]{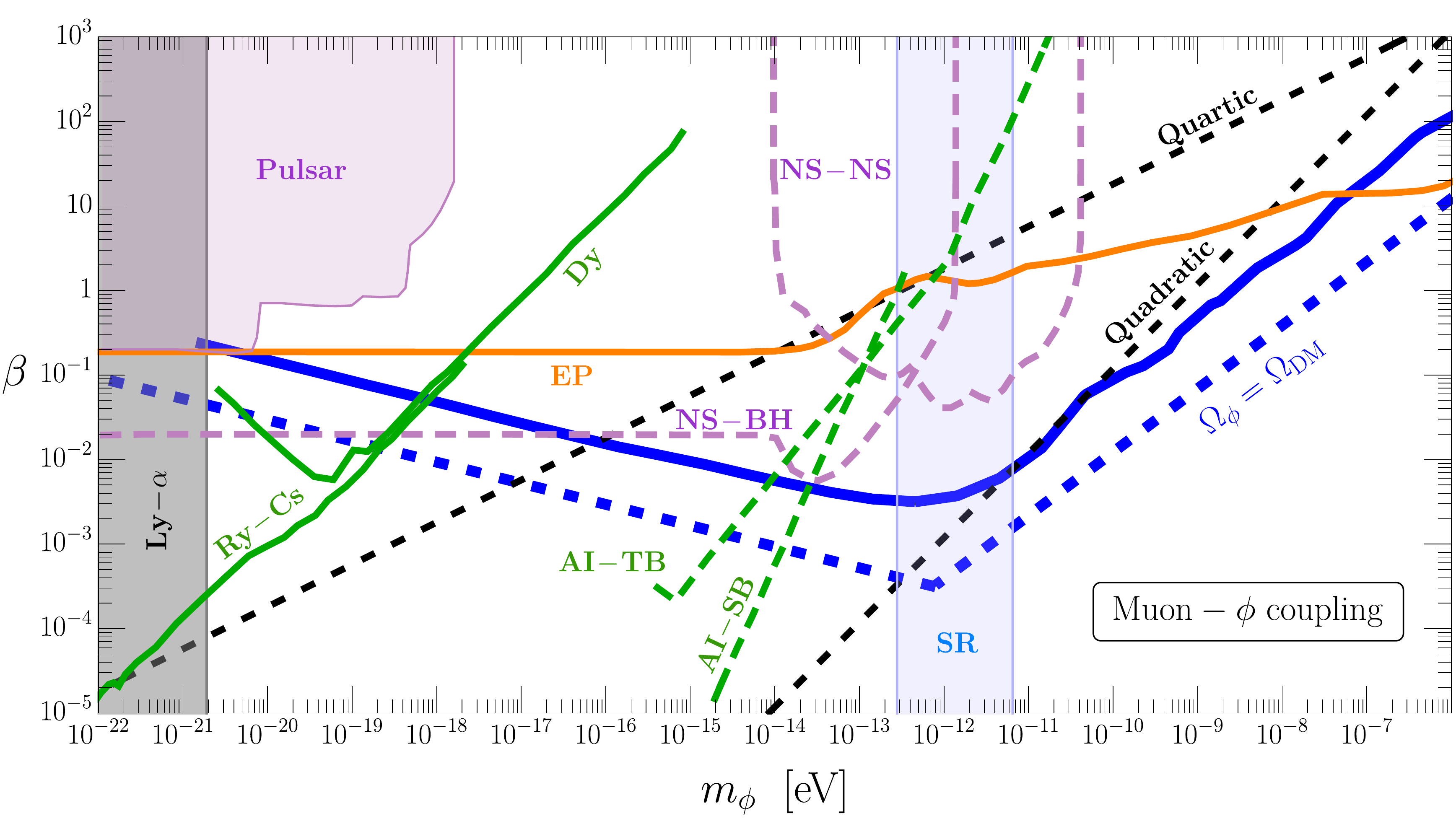}
\caption{Existing constraints and future prospects on a muon-$\phi$ interaction represented in the $m_\phi-\beta$ plane. The thermal misalignment production mechanism predicts the correct dark matter abundance, $\Omega_{\phi} = \Omega_{\rm DM}$, over a wide range of scalar masses, shown as dotted (solid) blue lines for the analytical approximation (exact numerical solution). The difference between the exact solution and analytical approximation comes from slightly overestimating $\phi(y_{\rm osc})$ in the analytical approximation as well as including the temperature variation in $g_{*}$ in the numerical solution. Further information on the experimental and observational constraints shown in the figure is provided in the main text.
\label{fig:muon}
}
\end{figure*}

{\bf Scalar dark matter coupled to the muon.}---
We now describe the phenomenology of a scenario in which $\phi$ couples to the muon, i.e., $\psi \rightarrow \mu$. To be consistent with the SM gauge symmetries, the required low-energy $\phi \overline \mu \mu$ coupling must emerge from the dimension-5 operator $\phi \, \overline L_L H \mu_R+{\rm h.c.}$, which may arise in a variety of UV completions above the weak scale; see for example Refs.~\cite{Batell:2017kty,Batell:2021xsi}. 
In Fig.~\ref{fig:muon}, we show the analytical (exact numerical) relic density target for this model with dashed (solid) blue lines along with the associated constraints and prospects. It is evident from Fig.~\ref{fig:muon} that a rich variety of experimental and observational probes are present in this scenario, as we now discuss.

The first class of probes rely only on the gravitational interactions of $\phi$. In particular, in the ultra-low mass ``Fuzzy DM'' regime \cite{Hu:2000ke,Hui:2016ltb} observations of the Lyman-$\alpha$ forest flux power spectrum lead to the bound $m_\phi\gtrsim 2 \times 10^{-21} $ eV~\cite{Irsic:2017yje}. 
Additionally, the existence of light scalars implies spin-down of rotating black holes (BH) through superradiance (SR)~\cite{Arvanitaki:2010sy}.
Observations of fast-spinning stellar-mass BHs in X-ray binaries therefore lead to constraints on the scalar mass~\cite{Baryakhtar:2020gao}, as shown in Fig.~\ref{fig:muon}.

There are also direct probes the $\phi \overline \mu\mu$ Yukawa coupling that generates scalar misalignment and controls the DM abundance.
Muons are naturally  present in neutron stars (NS), and the resulting radiation of the light scalar can lead to anomalous decay of orbital periods in pulsar binary systems as well as NS-NS and NS-BH mergers observed by gravitational wave detectors \cite{Dror:2019uea}. The existing constraints from pulsars and future constraints from NS mergers are shown in purple in Fig.~\ref{fig:muon}. We note that muon-storage ring experiments can provide additional direct tests of the scalar-muon coupling, albeit at larger values of $\beta$ above the cosmologically favored region~\cite{Janish:2020knz}. 

The $\phi \overline \mu \mu$ interaction radiatively induces an effective scalar-photon coupling $g_{\phi\gamma\gamma} \phi F_{\mu\nu} F^{\mu\nu}$. In the absence of additional UV contributions to this operator, the effective coupling is given by $g_{\phi\gamma\gamma} = -(\alpha \beta)/(6 \pi M_{\rm pl})$, where $\alpha$ is the fine structure constant. Such a coupling induces a long-range Yukawa force between matter that violates the equivalence principle (EP) \cite{Damour:2010rp}. The associated constraints from tests of equivalence principle \cite{Schlamminger:2007ht} are shown in orange Fig.~\ref{fig:muon}.
Furthermore, in the oscillating DM background, such a coupling leads to temporal variations in the fine structure constant, which can be probed by atomic clocks with Dysprosium (Dy)~\cite{VanTilburg:2015oza} and Rubidium and Caesium (Ry-Cs)~\cite{Hees:2016gop}, as well as future terrestrial (space) based atomic interferometer experiments~\cite{Arvanitaki:2016fyj} AI-TB (AI-SB). These are shown in green in Fig.~\ref{fig:muon}.

Finally, we note that the coupling of the scalar to muons results in quadratic corrections to the scalar mass as well as quartic corrections to the scalar potential, the latter of which prevent the scalar from oscillating like matter. These corrections are naturally small for small $\beta$ and large $m_\phi$ (below the dashed black-lines in Fig.~\ref{fig:muon}), while for large $\beta$ and small $m_\phi$ we assume they are fine-tuned away.
Further we note that at two loops the QED coupling of the muon to the photon also induces a scalar thermal potential of parametric size $\delta V_{T,\gamma} \sim - \alpha^2 T^4 \times  (\beta \phi/M_{\rm pl})$~\cite{Anisimov:2000wx}. For $T \lesssim m_{\mu}/\alpha$, $(y \lesssim 1/\alpha)$, the muon one loop contribution (\ref{eq:Veff1}) dominates. In particular, provided this condition is satisfied by the oscillation temperature $y_{\rm osc}$ given in Eq.~(\ref{eq:yosc}), the two-loop effect can be neglected. In Regions 1 and 2, $y_{\rm osc} \simeq \sqrt{\kappa/3\gamma}$, implying that for $m_{\phi} \lesssim \gamma m_\mu^2/ \alpha^2 M_{\rm pl} \sim 10^{-6}\, \rm{eV}$ -- the entire mass range studied here -- the one-loop effective potential (\ref{eq:Veff1}) controls the $\phi$ abundance. 

{\bf Conclusions.}---In this {\it Letter} we have presented {\it thermal misalignment}, a novel paradigm  for the cosmological production of ultra-light scalar DM.  Due to a tiny Planck-suppressed coupling to a fermion in the thermal bath in the early universe, the scalar field evolves towards the minimum of its thermal potential at large field values, generating large misalignment prior to the onset of oscillations. 
Unlike standard misalignment, thermal misalignment provides a regulating mechanism such that any scalar field initial condition respecting $\phi_{i} \ll \phi(y_{\rm osc})$ leads to the same relic density today, providing a precise prediction of the DM abundance in terms of the scalar-fermion coupling $\beta$ and the scalar mass $m_{\phi}$. 

If the fermions are muons, there is still viable parameter space for the scalar to be DM. Were any future experiments to detect such a scalar, the exact relation between $\beta$ and $m_{\phi}$ will be a strong smoking gun signal of our model. Alternatively, assuming standard cosmology our relic density line in Fig.~\ref{fig:muon} (solid blue) presents the strongest bound over much of the natural $(m_{\phi},\beta)$ parameter space. 

Avenues for future exploration of this paradigm are rich, including investigation of the Higgs portal as a UV completion of scalar-fermion couplings, consideration of higher dimension operators or couplings to different SM fields, and finite-temperature dynamics of pseudoscalar fields such as axion-like particles. 
We leave these possibilities for future work.

{\bf Acknowledgements.}---We thank John Lee and Hiren Patel for helpful conversations and Andrew Long for detailed comments on the manuscript.
The work of B.B. and A.G. is supported by the U.S. Department of Energy under grant No. DE–SC0007914.

\bibliographystyle{apsrev4-1}
\bibliography{rcdm}% Produces the bibliography via BibTeX.

\end{document}